\documentclass[12pt]{article}

\usepackage{amssymb} 
\usepackage{amsmath}
\usepackage{makeidx}
\usepackage{graphicx}

\def\BEn{\begin{enumerate}}
\def\EEn{\end{enumerate}}

\def\fro{\kern-2pt\leftarrow\kern-2pt}
\def\breve{\mathaccent"7014 }

\def\0#1{{\mathrm{#1}}}
\def\1#1{{\mathbb{#1}}}
\def\2#1{{\mathbf{#1}}}
\def\3#1{{\mathcal {#1}}}
\def\4#1{{\mathsf{#1}}}
\def\5#1{{\mathfrak{#1}}}
\def\6#1{\overline{#1}}
\def\7#1{{\check{#1}}}
\def\8#1{{\widehat{#1}}}
\def\9#1{{\breve{#1}}}

\def\dag{\dagger}
\def\adj{{^{\dag}}}

\def\pd{\partial}

\def\<{\langle}

\def\>{\rangle}

\def\Cl{\mathop{{\mathrm {Cl}}}\nolimits} %Clifford algebra
\def\ox{\otimes}

\def\x{\times}

\def\BE{\begin{equation}}
\def\EE{\end{equation}}
\def\BEA{\begin{eqnarray}}
\def\EEA{\end{eqnarray}}

\def\rar{\rightarrow}

\begin{document}

\title{\huge Finite Quantum Dynamics}
\author { 
David Ritz Finkelstein%
\thanks
{ School of Physics, 
Georgia Institute of Technology, Atlanta GA. df4@mail.gatech.edu}
\and 
Mohsen Shiri-Garakani%
\thanks
{Department of Chemistry and Physical Sciences,
Pace University,
Pleasantville NY. mshirigarakani@pace.edu}
}

\maketitle

\abstract{
We general-quantize the dynamics of the quantum
harmonic oscillator to
obtain a covariant finite quantum dynamics in a finite quantum time.
The usual 
central (``superselected'')  time
results from a self-organization.
Unitarity necessarily fails,  imperceptibly  for middle times 
and grossly near the beginning and end of time.
Time and energy interconvert
during space-time decondensation or melt-down,
at a rate governed by a constant like the Planck power.
}

\newpage
\section{Singular theories, singular  results}

A group must be semisimple
to be stable against experimental error (regular, robust, generic)  [Segal, Vilela].
We suppose that special relativity, general relativity and quantum theory 
are transitory phases
in the evolution of physical groups from singular ones based on axioms to  stable
ones based on experiment,  through variations
on one underlying regularization process,
{\em general quantization} ---  introducing small generic non-commutativities that
convert a non-semisimple  algebra
 into a simple one.

To unify quantum theory and general relativity
some  hold  one of the two theories  fixed
and adapt the other; we have 
made such efforts ourselves.
But there are deep errors in both theories 
that neither correct.
Now we take
 a more general perspective that embraces both quantization
 and special relativization
as limiting cases.

From this viewpoint the 
three main evolutions of physics in the early twentieth century 
have 
suggestive similarities.
\begin{enumerate}
\item Each introduced a  small new 
non-commutativity.

\item The scale of each modification is
set by a fundamental constant,
so small  that the predicted non-commutativity is indetectable
in older, coarser experiments.

\item Each changed a singular algebra to a more generic one.
\end{enumerate}

Thus boosts cease  to commute  in special relativity,
parallel transports in general relativity,
and filtrations in quantum theory.
The respective fundamental constants are $1/c^2$, $G$, 
and $\hbar$\/.
The algebras introduced are those of the Lorentz group, the Einstein group,
and the unitary group.

The three evolutions seem to 
be instances of one general process that we call  flexion,
and  when it increases the non-commutativity
of dynamical variables,  {\em  general-quantization}\/.
A  general-quantization  is a homotopy
connecting from a singular algebra to one that is
less singular, more generic, by
introducing a growing non-commutativity, a
 generalized curvature.
  The concept if not the term is due to Segal \cite{SEGAL} and
  is perhaps implicit in the work of In\"on\"u and Wigner
 \cite{Inonu1952} and  Flato \cite{FLATO}\/, and 
 is first made explicit and applied  in the work
of Vilela Mendes
 \cite{VILELA}.
Segal suggested that further quantization would not only stabilize the theory but might 
improve its agreement with experiment and cannot hurt it.
Varela began that process and
we explore it here.

Our search for the quantum of time started
from the broad idea that discreteness and continuity
are reconciled within a quantum theory \cite{FINKELSTEIN1969}.
This was not an adequate guide,
and it required problematic changes in the standard group structure that  seemed apt 
to 
conflict with experiment.
Segal's principle focuses the search for quantum time and 
transforms this problem into a solution.
The quantum theory must have a semisimple group to be
stable (generic), 
and any group is infinitesimally close to a semisimple one.
Therefore a change 
in the commutation relations of an existing unstable theory
is desirable,
and a sufficiently small change
suffices to stabilize the group,
accord with past data, and
make critical  predictions for future data.

A semisimple group  that provides stability
also permits  a finite-dimensional representation,
with bounded as well as discrete spectra for all operators, including time.
Full stabilization can be expected to eliminate all infinities.
Theories that are singular in having non-generic groups are 
also singular in making infinite predictions.
We need not distinguish between these usages of the term singularity.
Singular groups are not only unstable, they make physics blow up.

To deal with anti-commutation relations as well as commutation relations,
we generalize the simplicity principle from
Lie algebra to graded Lie algebra:
 \begin{center}
 \fbox{Graded commutator algebras of isolated physical systems are simple.}
 \end{center}
 This
leads back to Clifford-Wilczek statistics,
a general-quantized fermionic statistics already introduced by Wilczek
based on Clifford rather than Grassmann algebra
 \cite{WILCZEK}\/.

General quantization usually introduces new regularization operators 
or {\em regulators}
for some  commutators 
that are 0 in the singular limit,
and new regularization constants or {\em regulants}
for the final values of homotopy parameters.
Special relativization and canonical quantization
are exceptional. They
have regulants $c$ and $\hbar$ and  no regulators.
For general relativity, the  curvature tensor is a regulator and the gravitational constant $G$ is its regulant.

All these theories preserved or introduced some singularities 
that must now be regularized.
All the singularities of present physics can be traced to their source in
various singular Lie algebras.
L. H. Thomas, H. P. Snyder,  W. de Sitter, and I. E.  Segal
performed famous quantizations
that reduce these singularities
but
 still do not eliminate them, not
even when  are all combined and applied at once.
To make
the present physical algebras  semisimple  takes a  quantization
engineered for that purpose.

 While simplicity seems like a special case of semisimplicity,
in quantum theory  one  measurement
reduces the semisimple algebra to one simple subalgebra.  
So we shall
require simplicity without losing generality.

The simplicity principle implies much of quantum theory.
It excludes  classical mechanics, 
since the canonical group is not a Lie group.
The  simple Lie group that replaces the canonical group
 is the connected isometry group of some quadratic
space up to isomorphism, according to Cartan.
We may use this space as the state-vector space of the system,
and its quadratic form as the scalar product of the quantum theory.

The simplicity principle is suggested by the fact that the evolution of physical theories is at least
partly Darwinian.
Other things being equal --- which is not always the case --- 
stable theories are  better adapted 
to survive small improvements in measurements.
Finiteness is then
a somewhat unexpected reward for stability.
Present divergent  theories are but singular limits of 
finite simple generic ones.
We can construct  candidate simple theories by quantizing
the singular theories that work best.

Theories that are proposed as fundamental are usually selected to be 
special or distinguished as
opposed to generic,
for reasons of apparent simplicity.
For example, Euclid might have preferred his plane geometry to spherical  geometry 
on the grounds 
that a  flat space is ``simpler''  than a  curved one.
String theory and gauge theory  in their original forms, for example,
are set in commutative spaces and so
preserve one of the main  instabilities of the present quantum field theory. 
Gauge groups have arbitrary functions on space-time as group parameters 
and are not Lie groups.   They must be general-quantized 
to make them Lie.

Deformation quantization  \cite{FLATO}  uses a homotopy too 
but not the simplicity principle.
The canonical quantum oscillator,
with
all its singularities, seems to be an acceptable end point
for deformation quantization,
while it is only the starting point of Segal's  quantization.
The theory resulting from deformation quantization
 seems to have both
a singular algebra product and a more generic one
while nature and
the general-quantized theory seem to have
but one product.

The canonical Lie algebra $d\0H_1(x, \partial_x, 1)$
 of the differential calculus,
quantum theory
and gauge theory is unstable, singular, non-generic.
Therefore it
is probably just a transient phase
that we will  outgrow.
Segal  stabilized
it
by supplementing $\hbar$
with two new quantum constants $\hbar'$\/, $\hbar''$
of other dimensions.
This produced a  simple group
that has
the canonical algebra as a singular  limiting case,
has
continuous symmetries that are nearly  the
same  in a limited experimental domain
and
that can nevertheless become as large as we like, and is
nevertheless finite in volume and dimension; 
as the round Earth is practically flat in a limited domain,
yet finite.
One might say that we have just fully grasped how round the world
really is.

  Unlike lattice regularization,
general quantization does not reduce the relativity group of the space-time  to a discrete subgroup.
It merely changes the group slightly,
ultimately to a simple Lie group.
Snyder's quantum space-time and de Sitter's curved space-time
were limited forms of  general quantization.

Looking at physics  past, In\"on\"u and Wigner  
used a special case of 
Segal's general concept
to relate special relativity to Galileo relativity.
They formulated the concept of group contraction 
  \cite{Inonu1952},
a direct process, as
a linear transformation of the Lie algebra
that flattens it in the limit.
Contraction destabilizes the theory.
The Snyder and de Sitter general-quantizations were inverse contractions.
The quantization of quantum theory suggested by Segal 
 is not an inverse contraction, 
being necessarily non-linear.

The simplicity principle was in general circulation
by the 1960's.
Peter Bergmann mentioned it in a lecture, for example.
And some form of Segal's
 quantization  of the time-independent harmonic oscillator 
 is now under study by several groups from several  points of view
 \cite{Atakishiyev2003, Carlen2001, KUZMICH, MANFREDI, THOOFT}.
But the first work to take theory-stability seriously is that of Vilela Mendes.
 
 Looking to physics future, such reforms, pushed to their limit,
lead not only to a stable theory but also
 to
a finite one in a finite quantum space-time.
These have been sought by some physicists 
since the formulation of quantum theory.
 
 To stabilize present-day instabilities
requires several radical changes at once.
Present theories are {\em pointed} --- have absolute space-time points ---
and {\em local}  ---
couple these points only 
to their infinitesimal neighbors.
The Einstein group (of a manifold) ---like any other gauge group ---
is not simple precisely because it
respects points,
coupling $x^{\mu}$ into $\partial_{\mu}$ but not
conversely.
Such non-reciprocity is an infallible sign of a compound group.
When we make the Einstein group simple and generic,
we lose the space-time points.
A simple physics that reduces to a gauge theory in a singular limit 
must be not merely non-local
but
non-pointed.

More generally,
any local theory is singular and unstable, but an arbitrarily small quantization suffices
to make it simple and stable. 
It is then non-local and non-pointed.

Planck introduced the constant $\hbar$
in a way that
froze out the very stiff oscillators
responsible for the infinite heat capacity of cavity radiation
in Maxwell's theory.
Einstein recognized this effect
as a consequence of the 
quantum of radiation, the photon.
The canonical commutation relations 
and complementarity provided the conceptual framework
for this otherwise  mysterious regularization.
They  left
the zero-point energy of the resulting quantum theory 
of electromagnetism still divergent.
To be sure, Heisenberg later
replaced the local Lagrangian and Hamiltonian of Maxwell
by a non-local reordering 
that was arbitrarily tailored to have  zero ground-state energy-density.
This zero is not a prediction of the theory but an arbitrary assumption.
One expects any field theory to contribute to 
the zero-point energy of the vacuum, and so to a dark energy and mass,
but  present singular theories like quantum electrodynamics cannot 
predict this contribution.

The quantum theory of the linear harmonic oscillator, 
a constituent of all present quantum field theories, 
carries the seeds of some of
the divergences of quantum field theory.
Almost
all the  operators of this theory are unbounded, including its fundamental observables
$q, p$.
The basic operators of position $q$\/, momentum $p$\/, 
and  Hamiltonian $H\sim \frac 12 (p^2+ q^2)$
diverge on almost every vector $\psi$ in its Hilbert space:
$q\psi=p\psi=H\psi=\infty$. 
Since Von Neumann taught how to 
extract finite answers from such a
mathematical mine-field
we have become inured to
life on the brink of infinity.
But such infinities do not result from experiment but
from cosmological assumptions that go
beyond  experiment to
 ideology.
They  occur in a quantum theory if and only if
its Hilbert space is infinite-dimensional.
Finite-dimensional Hilbert spaces cannot 
represent singular algebras well;
but they can represent simple ones,
and so  simple or generic groups
makes a finite theory possible
with no loss of continuous symmetry.

By the {\em kinematical group} of a system we mean 
the group of all possible reversible actions on the system.
We replace many of the present principles of quantum theory by
the Segal simplicity condition, which excludes classical mechanics and classical field theories:

{\em The kinematical group  of an isolated physical system is simple.}

By a  finite or simple quantum theory we mean one obeying this simplicity condition.
Each simple Lie algebra is the  Lie algebra of the isometry group of
a unique 
finite-dimensional quadratic space;
this is the state-vector space of the system.
In a simple quantum theory all observables have finite bounded spectra.
Infinite-dimensional algebras can still be entertained, namely as singular limits,
for mathematical convenience;
just as the differential calculus is sometimes a convenient approximation to the calculus of finite differences.
But the symmetries of nature should already appear in the simple theory, which is the more physical theory,
and not only in its singular limit, which is less physical.

General quantization  does not make a theory that satisfies us aesthetically. 
We are all habitual transgressors of Ockham's Law.
(``Thou shalt not multiply entities unnecessarily.'')
If three data bars will admit a straight line,  we prefer the infinite line
to the many large circles that also pass through the data bars, though some circle 
almost certainly fits the data  better,
and is shorter.
This multiplies points unnecessarily.
Likewise, 
from our finite experience with many space-time points
we postulate that the set of space-time points is infinite
both in the large and in the small, again
breaking Ockham's Law. 
Segal's reform reaffirms Ockham's Law.
Somehow a large numerical constant offends our sensibility
when $\infty$ does not.

We test the simplicity  principle here on the
stationary (time-independent) and the dynamical (time-dependent)
 linear harmonic oscillator.
 
General-quantizing the stationary oscillator  introduces
two  post-quantum Segal constants $\hbar', \hbar''$ besides
the usual Planck constant $\hbar$\/.
The regularized coordinate and momentum now 
transform as infinitesimal rotations in SO(3).
We study how
the resulting finite quantum oscillator approaches the usual singular 
quantum oscillator as a singular limit.

General quantization freezes out the offending zero-point oscillations 
of extremely hard or soft oscillators
without
greatly changing the zero-point energies of medium ones.
The system Hilbert space becomes finite-dimensional and
its operations 
can in principle always be carried out.
The frozen oscillators  also grossly violate
the usual equipartition and uncertainty relations.
Algebra regularization clearly
 has profound consequences for extreme energy physics:
the physics of both very high and very low energies.

This toy model illustrates how a finite quantum theory of the cavity
can produce a finite zero-point energy without conflicting with
the many finite predictions and symmetries of the usual quantum theory.
We propose that  the  linear harmonic
field oscillators considered fundamental in present quantum physics 
-- we mean  those of allegedly fundamental fields, 
not those of crystals, say --- 
are actually 
dipole rotators in a three-dimensional space,
with fixed high angular-momentum quantum number
$l$ and with  third  angular-momentum component
$m\sim l$.
The unobserved oscillators responsible for 
the infrared and ultraviolet divergencies
of present quantum theories
are frozen by finite quantum effects described here
and contribute negligibly to the zero-point energy.

We then general-quantize the quantum oscillator dynamics. 
This leads to a  finite quantum theory of the 
dynamical harmonic oscillator in quantum time.

In dynamics 
the oscillator or field variable is a function of time.
When we call a theory c/c, q/c, or q/q,  the 
 denominator tells whether the
independent temporal  or T variable is c or q, and the numerator 
tells about the dependent   or S variable.
Quantum chromodynamics is a q/c theory.
Its field-variable system is quantum, with a non-commutative logic,
and the space-time is c, with a commutative logic.
The commutative logic leads to a singular algebra
and requires flexing.
Here we quantize the quantum oscillator q/c dynamics
and arrive at a q/q one
that is stable and finite.

\section{Regularization by quantization}

The totality of Lie products $\4x: \3A\ox \3A\to \3A$ admitted by a given vector space ${\3A}$,
also called structure tensors,
 form
a quadratic sub-manifold $\3J(\3A)=\{\4x\}$ 
of the tensor space ${\3A}\ox [{\3A}\adj\ox {\3A} \adj]$
(here the $\dag$ dualizes and the brackets skew-symmetrize),
defined by the Jacobi law, which is quadratic in $\4x$.
A  {\it regular} (stable, robust) algebra is a Lie algebra
that is unchanged  up to
isomorphism by all sufficiently small changes
 in its structure tensor (Lie product) within the manifold $\{\,\4x\,\}$.
For example, 
the Lorentz algebra is stable against small corrections to the speed of
light. 
By {\it flexion} we mean a homotopy of the structure tensor
of a compound algebra that it less commutative,
closer to semisimple.
When this makes the dynamical variables non-commutative it becomes quantization.
{\it Flattening} is  the inverse process.

\subsection{Flexing the canonical commutation relations}

The canonical (or Heisenberg) Lie algebra $d\0H(1)$ is
defined by the  canonical commutation relations
\BE
 \label{eq:Heisenberg2} 
p\,\4x\,q=-i\hbar  1, \quad
1\,\4x\,q=0,\quad
q\,\4x\,1=0.
\EE
among its three hermitian generators $q, p, 1$.
It is compound and 
the central unit $1$ 
generates its radical.
$1$ is an idol of the theory in the sense of Bacon \cite{BACON}.
Segal proposed to simplify $d\0H(1)$ by introducing 
a third variable $r$ to replace $1$, 
and two more quantum scale
constants, which we designate here by  
$\hbar'\equiv \hbar^{[1]}$ and $\hbar''\equiv \hbar^{[2]}$\/.
We also switch from hermitian observables $p, q, 1$
to anti-hermitian generators $\8p, \8q, \8r$.
Segal's
general-quantized commutation
relations are, except for notation,
\BE
\label{eq:SEGAL-}
{\8q}\,\4x\, {\8p}={\hbar}\8r,\quad
\8r\,\4x\, {\8q}={\hbar}' {\8p},\quad
\8p\,\4x\, {\8r}=-{\hbar}''{\8q},\quad
\EE
  \cite{Carlen2001,
VILELA,SEGAL}
For any $\hbar, \ \hbar',\ \hbar'' > 0 $ these relations
define the Lie algebra  $d\0S\0O(2, 1)$.
The irreducible unitary representations of this non-compact 
group are
infinite-dimensional. 
 Ultimately we will need an indefinite metric for relativistic reasons,
but not for the time-independent harmonic oscillator.
We therefore drop the minus sign and adopt the general-quantization
\BE
\label{eq:SEGAL}
{\8q}\,\4x\, {\8p}={\hbar}\8r,\quad
\8r\,\4x\, {\8q}={\hbar}' {\8p},\quad
\8p\,\4x\, {\8r}={\hbar}''{\8q},\quad
\EE
with constants $\hbar, \hbar', \hbar''
>0$ and group $\0S\0O(3)$\/.

Let us rewrite $q\equiv q^{[0]}$\/, $p\equiv q'\equiv q^{[1]}$\/,
$o\equiv q''\equiv q^{[2]}$ and
assume an invariant  Euclidean metric $g_{ij}$.
Then
 \cite{Baugh2003,Galiaudtinov2002}
\BE
\8q^{[i]}\,\4x\,\8q^{[j]}=-i\sum_k\epsilon^{ij}{}_k \hbar^{[k]} \8q^{[k]}.
\label{eq:SHeisenberg}
 \EE
We call $\8p,\  \8q,\  \8r$  momentum, position, and action generators,
respectively.  In this toy, $\8r$ is the sole regulator.

This general-quantized algebra looks as if 
 one has replaced  the imaginary quantum constant
 $\hbar i$ by a dynamical variable $\8r$, 
a process  one might call ``$ i$ activation.''
The relations among $\8p, \8q, \8r$ are symmetric under $\0S\0O(3, \1R)$.
Where the canonical theory has an absolute relation of canonical conjugacy
expressed by $[p, q]=-i\hbar$,
the general-quantized theory
has a relative relation of {\em canonical conjugacy with respect to} $r$\/,
expressed by  $[\8p, \8q]=\hbar \8r$\/.
For example,
 the  canonical conjugate  of $\8q$ with respect to $\8p$ is  $-\8r$.

In a previous exploration in quaternion quantum theory  
an activated  $\hbar i$  served as Higgs field
defining the electromagnetic axis $\eta(x)$ 
that resolves the
electroweak gauge boson into electromagnetic and weak
bosons \cite{FINKELSTEIN1963},
and
gives mass to the charged
partner of the photon through
the St\"{u}ckelberg-Higgs effect.
This led to a natural $\0S\0U(2)$ that was interpreted
as isospin. 
The activated $i\hbar$  generated rotations about the
electric (or electromagnetic) axis in isospin space, defining a
natural Higgs field. 
The quaternion  theory was dropped because it did not 
shed light on
color $\0S\0U(3)$.

Now we activate $\hbar i$ on the more
principled grounds of  Segal:  simplicity.
We expect that the third axis  will again give rise to
a Higgs field.
There is now
plenty of room for internal groups like color $\0S\0U(3)$,
though we do not seek them for the harmonic oscillator.

Flexing generally faces the same kind of factor-ordering problems
as quantization and gauging.  

Except for scale factors the simplified commutation relations are
those  of an $\0S\0O(3)$ quantum
angular-momentum operator-valued vector 
${\2L}=i{\2L}\,\4x\,
{\2L}$ for a dipole rotator in three-dimensional space. 
We assume an
irreducible representation with 
\BE 
{\2L}^2=l(l+1) 
\EE 
where $l$ can have any non-negative half-integer eigenvalue.
(In the present work it suffices to consider only integer values of $l$.)
Then 
$L_1, L_2,  L_3$  are 
represented by $(2l+1)\times (2l+1)$ matrices obeying 
\BE
{L}^i\,\4x\,{L}^j=-i \epsilon^{ijk}{L}_k\/.
\label{eq:LCR}
\EE
We fix the scale factors by setting
  \BE
q^{[k]}=\4Q^{[k]} L^{[k]} \quad \mbox{(No sum over $k$.)}
 \label{eq:scales} 
 \EE
 In the singular limit $l\to \infty$ and the oscillator
 is nearly polarized along the $L_3$ axis, with
 $L_3\approx l$\/.
By  (\ref{eq:LCR})  
\BE 
\4Q=\sqrt{\hbar\hbar'},\quad
\4Q'=\sqrt{\hbar \hbar ''}, \quad
\4Q''=\sqrt{\hbar'\hbar''} \;=\; 1/l\/.
\label{eq:QQ'Q''}
\EE
 The commutation relations (\ref{eq:LCR}) 
 and the angular momentum quantum number $l$
 determine a simple (associative) enveloping
algebra $\0A(\2L,l)$ of $(2l+1)\x (2l+1)$ matrices.
The
spectral spacing of  ${L}_3$ is 1, so the finite quantum
constants $\4Q, \4Q', \4Q''$ serve as quanta
   of the position, momentum and action variables.
Since $q, p$  have continuous spectra in quantum
theory, 
the constants $\4Q,  \4Q'$ must be very small on the ordinary quantum
scale.
It follows that $\4Q''=\4Q\4Q'/\hbar$ is also very small
on that scale and $l\gg 1$.

For $l\gg \sqrt{l}\gg 1$, 
variations $\delta (\8r^2) \le O(l^{-1/2}) \ll
1$ about $\8r^2=-1$ can be negligible at the same time as the
spectral intervals $\delta p\le  \4Q'\sqrt{l}$ and $\delta q\le
\4Q\sqrt{l}$ for  quasicontinuous $p, q\approx  0$.
This simulates the usual oscillator kinematics.

\section{Stationary harmonic oscillator}

\label{sec:STATIONARYOSCILLATOR}

In the section we general-quantize and regularize 
the time-independent linear harmonic oscillator
(\S\ref{sec:STATIONARYOSCILLATOR}),
recapitulating and  extending the work of \cite{SHIRI2005},
and then   the time-dependent one 
(\S\ref{sec:DYNAMICALOSCILLATOR}),
with additional differential operators $t$ and $\partial_t$.

General-quantizing the oscillator modifies the statistics.
The oscillator is the prototype bosonic aggregate.
Its Heisenberg algebra is that of the creators and annihilators of
structureless bosons.
When we change that algebra from that of an oscillator
to that of a rotator, 
we change the statistics from bosonic to one
that may be called finite-bosonic.
There is no bound on the boson occupation number, 
but
the finite boson has a finite bound $N$ on its occupation number. 
$N$ is one of the regulants of the regularized theory.
When $N\to \infty$, the finite-bosonic statistics approaches the
bosonic statistics of the singular theory.

A familiar ordering question arises at once.
Since general-quantizing introduces non-commutativity,
the  factor-ordering of some products is immaterial
in the flat theory but is significant in the general-quantized theory.
Re-ordering  the singular theory will
 introduce a correction of order $h'h''$
that may be experimentally significant.
This makes general-quantization as  notation-dependent
and ambiguous 
as  quantization and relativization.

The Hamiltonian of the general-quantized harmonic oscillator is
 \BE \label{eq:FQHO}
H=\frac{\4Q'^2}{2m}L^2_2+\frac{k\4Q^2}{2}L^2_1= :\frac K2\left ( L_2\/^2
+ \kappa^2 L_1\/^2\right)
 \EE 
 where
\BE
K:=\frac{(\4Q')^2}{m},\quad \kappa^2=\frac{\hbar'mk}{\hbar''}. 
\EE 
For fixed $\hbar^{[k]}$, all  finite oscillators  are divided
into three kinds with ill defined boundaries:
{\em medium}\/,  where kinetic and potential terms in $H$ are of
comparable size ($\kappa \sim 1$);  {\em soft}\/, when the potential energy
term is dominant ($\kappa\to 0 $); and hard, when the kinetic energy
term is dominant ($\kappa \to \infty$).
For a  scalar field
in  standard singular quantum field theory, 
the  oscillators that
give rise to infrared divergencies 
become soft oscillators in the finite quantum theory,
and those that
feed ultraviolet divergencies
become hard oscillators.

To represent the general-quantized time-independent algebra,
let $\gamma_{{\sigma}}(n)$ ($\sigma=1, 2, 3$\/, $n=1,\dots, N$) be $3N$ Clifford generators of positive signature.
Each $\gamma_{{\sigma}}(n) $ represents an elementary process that toggles the occupation number of one ``chronon'' --- quantum of space-time-etc.  --- with Clifford-Wilczek statistics.

Then we take 
\BEA
q &=&  \4Q \sum_n \gamma_{31}(n),\cr
p &=& \4P \sum_n \gamma_{23}(n),\cr
r&=& \4R \sum_n \gamma_{12}(n)
\EEA
as regularized anti-coordinate, anti-momentum, and anti-action of the oscillator at one instant of time.

Assume that $N$ is even. Then the spinors on which these operators act have $2^{3N/2}$ components before reduction of the matrix algebra into  irreducible representations.

\subsection{Medium oscillators}
The  case $\kappa=1$ 
is  symmetric under rotations about the $z$ axis, and so is especially simple
\cite{THOOFT}.
Since
\BE
 ({L}_1)^2+({L}_2)^2+({L}_3)^2=
{\2L}^2=l(l+1),
\EE
\BE
{H} = \frac{K}{2}\left(l(l+1)-({L}_3)
^2\right)\label{eq:Hamiltonian}
\EE
The oscillator quantum
number $n$ that labels the energy level is now
\BE 
n=l+m. 
\EE
The general-quantized energy spectrum
is
\BE E_n= \frac{K}{2}\left(l(l+1)-(n-l) ^2\right) = {lK}\left(n +
\frac{1}{2} -\frac{n^2}{2l}\right) \EE
For $n\ll\sqrt{l}\ll l$ this
reproduces the usual uniformly-spaced oscillator energy spectrum as closely as
desired, but with multiplicity  2 for each level instead of 1.

The ground-state energy for this oscillator is  
\BE E_0 =
\frac{1}{2}Kl=\frac 12  \hbar\omega,
\label{eq:groundenergy}\EE
exactly the usual
oscillator ground energy, 
since $Kl=\hbar\omega$.

The main new feature is that this finite oscillator has an upper energy limit 
\BE E_{max}= \frac{1}{2}Kl(l+1)\EE
 as required by  a finite quantum theory.

In the general case of $\kappa \sim 1$ we obtain
an upper bound for the ground energy by a variational
approximation with the  trial function $|L_3=\pm l\>$.
This 
reproduces our previous result (\ref{eq:groundenergy}),  now
as an upper bound for the ground energy of a
medium FLHO:
\BE E_0\leq \frac{1}{2}Kl.\EE
Medium oscillators
have many states with $m$-value close to the extremum values $m=\pm l$.
The usual
Heisenberg uncertainty principle 
\BE(\Delta p)^2(\Delta
q)^2\geqslant\frac{1}{4}\< i p\,\4x\,q\>^2=\frac{\hbar^2}{4}.
\EE
becomes
\BE
 (\Delta L_1)^2(\Delta L_2)^2\geqslant
\frac{\hbar^2}{4}\< L_3\>^2_{|L_3\approx\pm l\>} \EE 
for a low-lying energy level of a medium 
oscillator.
By (\ref{eq:scales}) and (\ref{eq:QQ'Q''}), 
 \BE
  (\Delta p)^2(\Delta y)^2\geqslant  \frac{\hbar^2}{4}\EE 
  for large $l$. 
  So 
medium oscillator states in low-lying energy levels have uncertainty products
consistent with the Heisenberg uncertainty principle.

\subsection{Soft  oscillators}

When $\kappa \ll 1$
we  can estimate the spectrum  of $ H$ by
perturbation theory.
The unperturbed Hamiltonian is
the kinetic energy  
\BE
H_0=\frac{K}{2}L^2_1\/.
 \EE 
The unperturbed eigenvectors are $|L_1=m\>
$\/.
The unperturbed energy levels are 
\BE
E_m(0)=\frac K2 m^2. 
\EE
The first-order shifts
are 
\BE \delta E_m=\frac K2 \< L_1=m| L^2_2|L_1=m\>. \EE 
Due
to the axial symmetry of $|L_1=m\>
$,
\BE \< L_1=m| L^2_2|L_1=m\> = \< L_1= m|
L^2_3|L_1=m\> . \EE 
Therefore the energy shift is
 \BEA
 \frac K2  \< L_1=m| \kappa^2 L^2_2|L_1=m\>&=&
 \frac K 4  \kappa^2 \<  m| L^2_1+L^2_2|m\> \cr
&=&\frac K 4 \kappa^2 \< m| L^2-L^2_3|m\>\cr
&=&\frac K 4 \kappa^2  l(l+1)-m^2 
\EEA 
  to lowest order in $\kappa^2$,
The general-quantized energy spectrum is then
\BEA
 E_m&\approx &\frac{K}{2}m^2+\Delta E_m\nonumber
\\&=&\frac{K}{2}m^2+ \frac{1}{4}K\
\kappa^2\left[l(l+1)-m^2\right]\label{eq:SoftFLHO}\EEA
The estimated upper
bound for the energy is
\BE
E_{max}\approx \frac{1}{2}Kl^2(1+\frac{\kappa^2}{2l})\EE

For $\kappa \rar 0$ this reproduces the upper bound for the
unperturbed Hamiltonian $L^2_3$, as it should.
The zero-point energy $E_0$ of first-order perturbation theory is
\BE E_0\approx \frac 14 \kappa^2 K l(l+1) \EE
For $\kappa \to 0$ this is infinitesimal compared to the standard quantum oscillator.

 A soft oscillator  shows little resemblance to the usual 
quantum oscillator.
Its  energy levels  do not have  uniform
spacing.
Its  kinetic energy 
dwarfs its potential energy,
grossly violating
equipartition.
 The low energy states
are near $|L_1=0\>$ instead of $|L_3=\pm l\>$.
Its  $p$
degree of freedom is frozen out. It is ``too
soft to oscillate:"
There is not enough energy in the $q$ degree of
freedom, even at its maximum excitation, to produce one quantum of $p$.
The uncertainty relation reads
 \BE
 (\Delta L_1)^2(\Delta L_2)^2\geqslant
\frac{\hbar^2}{4}\< L_3\>^2_{|L_1\approx 0\>}\approx 0 \EE
Therefore \BE \Delta p \Delta q \ll \frac{\hbar}2, \EE
which violates the Heisenberg uncertainty principle grossly.

\subsection{Hard  oscillators}

Hard  oscillators reverse the story but violate the same basic  principles 
of the caninical quantum theory as soft oscillators.
 A hard oscillator has much greater potential than kinetic energy.
Its low energy states are now near
$|L_2=0\>$ instead of $|L_3=\pm l\>$ (the medium case) or $|L_1=0\>$
(the soft case).
Its  $q$ degree of
freedom is frozen out. It is ``too hard to
oscillate.'' There is not enough energy in the $p$ degree of freedom,
even at maximum excitation, to arouse one quantum of $q$.

A hard oscillator can likewise be treated by perturbation methods. 
The
kinetic energy is the perturbation. 
We may carry all the of the main results in the previous section for
soft  oscillators to the hard ones simply by replacing $\kappa$
with $1/\kappa$ and $K$ with $K\kappa^2$.
A hard oscillator shows no resemblance to the
usual quantum oscillator.
Its zero-point energy $E_0$  is now
\BE E_0\approx \frac K4   l(l+1) \EE
For $\kappa \to \infty$ this is infinitesimal compared to the usual
quantum oscillator zero-point energy. 
Its energy levels of a hard oscillator are not
uniformly spaced. 
Its  uncertainty relation reads
 \BE
 (\Delta L_1)^2(\Delta L_2)^2\geqslant
\frac{\hbar^2}{4}\< L_3\>^2_{|L_2\approx 0\>}\approx 0 \EE
Therefore 
\BE \Delta p \Delta q \ll \frac{\hbar}2, \EE
which seriously violates the Heisenberg uncertainty principle again.

\subsection{Unitary Representations} %0050611

Variables $p$ and $q$ do not have finite-dimensional unitary
representations in classical and quantum physics. They are
continuous variables and generate unbounded translations of each other.
But since in the
general-quantized quantum theory
all operators become finite and quantized, we
expect all translations to become rotations,
 with simple
finite-dimensional unitary representations.

The canonical group of a classical oscillator becomes
the unitary group of an infinite-dimensional 
 Hilbert space for a quantum oscillator,
and the unitary group of a $2l+1$ dimensional Hilbert space
for the general-quantized quantum oscillator.

The Lie algebra generated by momentum and position 
as infinitesimal symmetry generators is 
$\2H(1)$ for the classical and quantum oscillator
and  the $\2S\2O(3)$ angular momentum algebra 
for the finite oscillator.
The corresponding Lie algebras are the Heisenberg Lie algebra  $d\0H(1)$
and the orthogonal-group Lie algebra $d\text{SO}(3)$.

The commutation relations 
${\2L}\,\4x\, {\2L}=-i{\2L}$
and the angular momentum
quantum number $l$ 
determine a simple matrix algebra
$\0A(\8{\2L},  l)$; 
here $l$
can be any non-negative half-integer.
The spectral spacing of
the operators ${L}_k$ is 1,
so
the  constants
$\hbar_{k}$ serve as quanta
 of the $\8q_k$
 respectively.
Since $q, p$
have continuous spectra
in the flattened quantum theory,
their quanta $\hbar_1, \hbar_2$
must be small on the ordinary
quantum scale $\hbar\sim 1$.
It follows that $\hbar_3$ is also small on that scale.

For $\sqrt{l}\gg 1$, variations $\delta ({\8r}^2) \le O(l^{-1/2}) \ll 1$ 
from ${\8r}^2=1$ can be negligible
at the same time as the spectral intervals $\delta p\le  \hbar_2\sqrt{l}$ and 
$\delta q\le \hbar_1\sqrt{l}$ 
for  quasicontinuous $p\ll \hbar_2, q\ll \hbar_1$.

In the canonical quantum theory, $q$-translation is a continuous 
one-parameter unitary
subgroup of the kinematical group,
but not in the simple quantum theory,
obviously.
If the infinitesimal advance of $q$ were represented by a hermition operator,
$q$ would have a continuous spectrum,
and the Hilbert space would be infinite dimensional,
and all values of $q$ would have the same spectral multiplicity.
In fact the Hilbert space is finite-dimensional
and each eigenvalue $\lambda=iq'$ of $iq$ has a multiplicity  $M(\lambda)$
that is maximum for $q^2=0$, varies slowly with $q^2$ for $q\sim 0$,
and goes to 0 linearly in $\lambda$\/.

\subsection{Quantum internal space}
\label{sec:INTERNAL}

It is  natural to ask what space 
the oscillator moves in.
In c physics we usually describe a system by a set of states, or state-set,
with some structure.

In quantum theory we may specify a simple system $\3T$ by an associated finite-dimensional 
state-vector space, that we designate by  $\0V \3T$,  or
by its algebra of coordinates $\0A \3T=\text{Endo V}\3T$.
The kinematics of the  q oscillator at one time is
defined by an infinite-dimensional irreducible representation of the 
three-dimensional complex Heisenberg algebra 
$d\text{H}(1)=\0A[q,p; \1C]$.

The general-quantized theory replaces
$\hbar i\in \1C$  by r $\hbar i \8r$, 
with regulator $r$\/.
We naturally define
the
corresponding  quantum  phase space
by the complex algebra
$\0A[\8q,\8 p, \8r; \1C]=d\0S\0O(3)$
with three generators $q \sim L_1, \; p\sim L_2, \; r\sim L_3$ isomorphic 
to the three components of a three-dimensional angular momentum vector $\2L=(L_1, L_2, L_3)$.
This small change in the commutation relations changes the spaces drastically
in the large.
The three variables $L_1, L_2, L_3$ are on the same footing in the regular theory
and are related to a central quantum number $l$ by
$(L_1)^2 +(L_2)^2 +(L_3)^2 =l(l+1)$. 
Now the irreducible representations are finite-dimensional, 
of dimension $2l+1$\/.

% This raises the charming possibility, still 
% speculative, that the still mysterious internal spaces of the standard model
% are vestiges of a similar quantization of space-time, near its classical limit.
% The fact that the internal variables commute with the space coordinates
% and momenta
% would then be no more surprising than the fact that $\8r$ 
% commutes with $q$ and $p$
% near the singular limit..

The regularized oscillator Hamiltonian $H=(p^2/2m) + (kq^2/2)$
  is that of a rigid dipole rotator with 
one infinite principle moment of inertia $I_3=\infty$
and with
a sharp total angular momentum quantum number $l<\infty$.
It has
a finite number of states
$\4N=2l+1\sim 1/(\hbar'\hbar'')$.
In the sinular limit $\4N\to \infty$\/, so $\4N$ is a regulant.
Manfredi and Salasnich discuss the 
statistical distribution of the levels  of the energy spectrum of the triaxial rotator 
and 
point out that part of the rotator spectrum 
can approximate the spectrum of a linear harmonic oscillator
\cite{MANFREDI}. 

Obviously the operator $\hbar \omega (L_12+ l))$ has
exactly the energy spectrum  of the singular theory, cut off
at the $\4N$th level. 
There is presumably a modification f $\8H'$ of $\8H$
that has exactly this equally spaced spectrum and
differs from our quadratic $\8H$ by 
corrections that vanish in the singular limit.
This equally spaced energy spectrum
eliminates the interactions between the quanta of excitation
when their number is less than $\4N$,
while the impossibility of higher occupation  than $\4N$
 can be regarded as an effective infinite repulsive $\4N+1$-body potential.

\subsection{Comparison of regular and singular quantum oscillators}

Let us compare the classical, quantum, and finite 
linear harmonic oscillators 
\cite{SHIRI2004}.

The general-quantized quantum oscillator is isomorphic to a dipole rotator 
with   Hamiltonian of the special form
\BE
H= \frac  12  K_x (L_x)^2 + K_y \frac 12   (L_y)^2, \quad 
K_x=\frac{\4P^2}{\mu}, \quad
K_y=\frac{\4Q^2}{\lambda}\/.
\EE
The classical and singular quantum 
oscillators have continuous coordinates and momenta.
The finite-oscillator
 position and momentum variables are quantized with
finite, uniformly spaced, spectra, with spacing $\4P, \4Q$ respectively,
and maximum values $l\4P, l\4Q$.
To pass for the more familiar singular ungeneral-quantized  oscillators of present-day physics,
a general-quantized oscillator must have many states,  $\4N=1/\4J\gg 1$.
But such oscillators are accompanied by some that have few states.

In the classical  theory all oscillators are isomorphic up to scale.
All singular quantum linear harmonic oscillators are likewise  isomorphic up to scale.
The constants $\hbar,\hbar''$ finally break this scale invariance.
The finite quantum linear harmonic oscillators
 fall into three
broad classes, which we term
{\em soft}\/, {\em medium}\/, and {\em hard}, 
according to the dimensionless ratio $K_y/K_x=\kappa^2$
of maximum possible potential energy  to maximum possible kinetic energy.

Medium oscillators ($\kappa \sim 1$ ) have $\sim \sqrt N$ low-lying states
with nearly the same zero-point energy and level spacing 
as the QLHO, like
rotators nearly polarized along 
the $z$ axis with $L_z\sim \pm l$.
They resemble the singular QLHO, obeying
the Heisenberg 
uncertainty principle and the equipartition principle 
when they are in their low-lying energy levels.

The soft and hard FLHO's  do not resemble the QLHO at all.
Their low-lying energy states correspond to rotators 
with  $\kappa\sim 0 $ or $\kappa\sim \infty$.
Their 0-point energy is infinitesimal compared to the QLHO.
They grossly violate both the uncertainty principle and equipartition 
in all their states.

Soft oscillators have frozen momentum 
$p\sim 0$,
their maximum potential energy being too small for even
one quantum of momentum.

Hard oscillators (kinetic $\ll$ potential) have frozen position $q\sim 0$,
their maximum kinetic energy
being too small for even one quantum of position.
These quantum freezings of degrees of freedom
resemble but extend 
the original  ones by which
Planck obtained a finite  thermal distribution of cavity radiation.
Even the 0-point energy of a similarly regularized field theory
will be finite, and can therefore be physical.

For the standard linear harmonic oscillator,
\BE
\8E-\8H=A L_{14} -BL_{23}^2 -CL_{24}^2
\EE
with real positive constants $A,B,C$. 
For medium oscillators this approaches the singular theory 
as $\hbar_{rp}, \hbar_{rq}\to 0$.
For soft oscillators the mass dominates the spring, $C\ll B$
and may be treated as a perturbation. The kinetic energy perturbation 
$CL_{24}^2$ happens to commute  with the unperturbed energies
$A L_{14} -BL_{23}^2$.

\section{Dynamical harmonic oscillator}
\label{sec:DYNAMICALOSCILLATOR}

Now let us
flex  the dynamical or time-dependent
oscillator.
staying as close to the singular theory as regularity
permits.
Above all we maintain and extend  the correspondence principle.  
The variables and equations
of the general-quantized theory converge (non-uniformly) to those of the singular theory
as  $\4T\to  0$ and $\4N \to \infty$.

This is a critical test of the  general-quantization   strategy.
We had not succeeded in making a reasonable q/q dynamical theory
before now.

\subsection{Forms of dynamics}
\label{sec:FORMS}

Let us designate the  q system under study at some one instant by $\3S$, 
and the system of c times over which we study it by $\3T$.
Here we general-quantize a 
q/c dynamical theory of $\3S$ over $\3T$ into a q/q theory.

The standard Hilbert-space structure is not enough to formulate a dynamical theory;
it must be supplemented by a theory of time. 
The canonical  dynamics takes the time axis to be $\2R$,
postulates a fixed Hamiltonian $H(t)$ possibly depending on $t$\/,
 and assumes a dynamical equation of the form
\BE
\label{eq:HEISENBERG}
i\hbar \frac{dq(t)}{dt}-H(t)\,\4x\, q(t)=[E-H] \,\4x\,q(t)=0\/.
\EE
This dynamics does not relate mere observables  like $q$ 
but entities like $q(t)$ of a separate category,
with separate meaning and structure.
We begin by giving a commutator algebra for  this larger structure.

The canonical algebra  has 
a complete set of commuting variables all associated with one time,
and the elements of the algebra are functions of only one time variable.
This is a single-time or {\em synchronic} theory.
Variables of a synchronic algebra are independent and grade-commute if they
are associated with spatially separated events.
Later variables, however, are not independent of early variables, but are
identified with combinations of them determined 
by integrating the canonical equations of motion.
This makes a non-relativistic distinction between space and time.

In the many-time or {\em diachronic }  form of dynamical algebra  
there are independent grade-commuting
variables at each space-time point, 
regardless of whether the separation is spacelike or timelike. 
The elements of the diachronic algebra represent functions  of space-time points.
The vectors on which they act describe histories; we call them
path vectors or history vectors.
The dynamical equations are now subsidiary conditions, not operator equations.
They single out 
a subspace of dynamically allowed path-vectors  $\Psi$ that  satisfy dynamical equations of the
vector form
$L\Psi=0$,  instead of equations of  the operator form $L=0$ of the synchronic theory.
where $L$ is some linear operator to be specified.

One can determine a synchronic state-vector by measurements all at one time. 
To determine a diachronic path-vector one must suspend the canonical  equations and make measurements at every time. 
Since we attribute the canonical equations to a condensation, this turning-off 
is not a purely mathematical fantasy. 
It may happen in  a change of phase of the ether.

A fixed Hamiltonian is built into structure of the synchronic algebra.
The diachronic  algebra is completely defined without reference to any Hamiltonian.

These considerations apply to both the c/c and  the q/c dynamics.

In this first study we general-quantize only the synchronic dynamics.
We suppose that a definite Hamiltonian is valid from input to output,
and use it to identify later variables with combinations of earlier ones. 
Any measurement overrides this Hamiltonian,
so this identification runs only between input and output times.

The q/c synchronic dynamics gives time two special roles:
\begin{itemize}
\item
 $t$ is a central operator, a superselection rule \cite{PIRON}\/.
 \item Experiment presents us with dynamics as a 
 system of unitary transformations $W(t, t')$ that
 connect any two values of time $t, t'$
 and the associated state-vectors $\psi(t), \psi(t')$.
 \end{itemize}
 Requiring all observables to commute with $t$  is not
generic.
 When $t$ is non-central,
 the dynamical correlation between different times
appears as an off-diagonal long-range order, off diagonal in $t$\/.
  Condensations create the central $q$'s and $p$'s of classical mechanics.
It is convenient to use the same language for central time.
 We treat central  time as if it resulted from a condensation
of a more generic dynamics.
 
When we combine systems, quantities may combine
in three useful ways:
\begin{enumerate}
\item Multiplicatively, like finite symmetry group operations;
\item Additively, like infinitesimal symmetries;
\item Identically, like time $t$ and $i$ in q/c theory
\cite{FINKELSTEIN1996}.
\end{enumerate}
We attribute the identical mode of composition to a 
widespread condensation that correlates time variables in many systems to
one another.
Underlying the identified variables of the condensate are additive variables
of the un-condensate.

The Heisenberg Lie algebras $d\0H(N)$ and groups
have at
 least two natural flexings,
the unitary and  the orthogonal line of groups.
We present them next and then choose one.

\subsection{The A line}
Baugh \cite{Baugh2004} regularizes the Heisenberg algebra $d\0H(n)$  within the unitary-group  Lie algebra $d\0S\0U(n+1)$.
He
introduces a high-dimensional linear algebra 
$d\mathrm{SL(n+1; \1C)}$ with generators $\Lambda^{\mu}{}_{\nu}$;
$\mu=0, 1,\dots, n$, with extra dimension 0.
The $n+1$ generators $\Lambda^{\mu}{}_{\mu}$ (no sum) are
related by $\Lambda^{\mu}{}_{\mu}=1$ (sum!).
\cite{Baugh2004}.
The Baugh regularization can be written by adding an index-value  0 and setting
\BE
q^{\mu}\fro \Lambda^{\mu}{}_0, \quad p_{\mu}\fro \Lambda^0_{\mu}\/,
r\fro \Lambda^0_0\/.
\EE
Its regulators are all the remaining $n^2$ independent generators $\Lambda^{\mu}_{\nu}$ ($\mu, \nu \ne 0$).

For a unitary representation of this abstract Lie algebra
 in a hermitian space
  with the usual positive-definite metric $\|\psi\|=\sum \psi^{\mu}{}^* \psi^{\mu}$
  it is convenient  to choose the Hermitian operators
\BE
q^{\mu}\fro \frac 12 [\Lambda^{\mu}{}_0+ \Lambda^0_{\mu}], \quad p_{\mu}\fro\frac i2 [\Lambda^{\mu}{}_0- \Lambda^0_{\mu}]\/.
\EE

\subsection{The D line}

Alternatively  one may present the $n$ $q$'s and $p$'s of $d \0H(n)$ 
as singular limits
of generators $o^{\mu\nu}$ of an orthogonal group $\0S\0O(n+2)$:
\BE
q^{\mu}\fro \4Q o^{\mu\,(n+1)}, \quad p^{\mu}\fro \4P o^{\mu\,(n+2)}, 
r\fro \4R o^{(n+1)(n+2)}\/.
\EE
We have no need of odd $n$; and
the even-$n$ groups lie on the D line.
This introduces two new index values $n+1, \; n+2$
(instead of only one for the A line)
and regulators $o^{(n+1)(n+2)}, \; 0^{\mu\nu}$ 
that must freeze out in the singular limit.  
It also introduces  regulants
$\4Q, \;\4P,\; \4R:=1/\4N$ that approach  $0$ or $\infty$ in the singular limit.
If the singular limit has the largest possible orthogonal group $\0S\0O(n)$ 
as symmetry group
then these
are the only regulants. 
If the limit reduces $\0S\0O(n)$ to the direct sum of 
$m$ smaller orthogonal groups, each of these  has its own trio of regulants 
$\4Q_i, \4P_i,
\4R_i$. 

Minkowski space-time in $n$ dimensions has an orthogonal group on the
D line;;
Hilbert space has a unitary group on the A line.
Which line shall we take? 
Special relativity suggests the D  line of groups
and quantum kinematics suggests the A line.

It would simplify matters if there were a clear hierarchic relation 
between special relativity and quantum theory, 
if one could say that quantum theory is the more fundamental, and so take the A line. 
But quantum theory imports its time variable from macroscopic relativity physics, 
and its quantum  imaginary is directly linked with time 
by its transformation under (Wigner) time reversal.
When we reverse the sign of time we reverse the sign of $i$\/.

It seems to be a useful general principle that what can transform can also change.
In the present discussion this suggests that $i$ is a variable like $t$; 
that the stable linearity of quantum theory is a real linearity, not a complex one.
The decentralization of time leads us to consider the decentralization of the associated $i$.
This happens naturally along the D line, which we tentatively follow.
The choice is moot for the oscillator, because
the D and A lines separate only beyond the group $\0S\0O(3,1)\sim \0S\0L(2,\1C)$\/
that we use for the general-quantized oscillator dynamics.
 
\subsection{Flexing the dynamics} %050424

\label{sec:FLEXALGEBRA}

Now we general-quantize the synchronic oscillator dynamics  
sketched in \S\ref{sec:FORMS}.
This requires us to express the q/c dynamics in the language of Lie algebra.

We replace  the anti-Hermitian generators 
$\7q:=iq, \ \7p:=ip$ of the time-independent oscillator algebra
by functions of time $\7q(t),\ \7p(t)$
obeying the canonical equation of motion.
To them and $i$ 
we adjoin  anti-Hermitian generators $\7t:=it$\/, $\7E:= \partial_t$\/, 
and the anti-(Hermitian) Hamiltonian $\7H:=iH$\/.
In the singular q/c theory $\7t$
commutes with $q,\ p, \ i$\/, and $\7H$,
and  $[\7t, \partial_t]=i$\/.
Let us write s $\partial_q,\ \partial_p$ 
for  Fr\'echet derivatives
with respect to the non-commuting variables $\7q, \ \7p$\/
\cite{FINKELSTEIN1955}.
We  define algebra elements
\BEA
\dot{ \7q }&:=& \frac{1}{\hbar} [\7H, \7q],\cr
\dot {\7p}&:=& \frac{1}{\hbar} [\7H, \7p],\cr
d/dt \equiv D_t &:= &\partial_t + \partial_{\7q} \cdot \dot {\7q}+
\partial_{\7p} \cdot \dot {\7p}
\EEA
Then the canonical dynamical equations 
\BE
\label{eq:CANONICAL}
[D_t, X] = [\partial_t, X] +\frac1{\hbar } [\7H, X].
\EE
are required to hold for every element of the synchronic algebra. 
 
In general $\7H$ is a given algebraic expression in $\7q, \7p, \7t$,
and  its commutators with them 
are generally not linear combinations of them.
Then we must adjoin commutators iteratively
 until the algebra closes under commutation. 
For the toy oscillator, with its quadratic Hamiltonian, this step is unnecessary.
The six generators $\7q,\ \7p, \ i, \ \7 t, \ \partial_t,\ \7H$ already close in
the q/c synchronic dynamical algebra $L_{\0D}$\/.

We drop  constants and accents  hereafter, writing $q,\dots, H$ for six general-quantized anti-symmetric generators. 
We use the prefix anti- to remind ourselves that these are anti-Hermitian.
Then the defining singular relations are
\BE
\begin{array}{rclcrclcrclcrclcrcl}
[q,p]&\sim & i,  &\ & [q, i]&=&0, &\ & [q, t]&=&0,&\ &[q,\partial_t] &=&0,&\ &
[q, H]&\sim & p,\cr
&&&&[p,i]]&=&0, &\ &
[p,t]&=&0,&\ &
[p,\partial_t]]&=&0,&\ &
[p,H]]&\sim&-q,\cr
&&&&&&&&
[i, t]]&=&0, &\ &
[i,\partial_t]&=&0,&\ &
[i, H]]&=&0,\cr
&&&&&&&&&&&&
[t,\partial_t]&=&-i, &\ &
[t, H]&=&0,\cr
&&&&&&&&&&&&&&&&
[\partial_t, H]&=&0.
\end{array}
\EE
The canonical equations (\ref{eq:CANONICAL}) are identities in virtue of these relations.

We general-quantize this algebra $L_{\0D}$ to $d \0S\0O(3,1)$, 
with state-vector space  $u:=4\1R$.
We
represent anti-time by $\8t = \4T L_{23}$,
anti-energy by $E=\4E  L_{24}$\/, oscillator anti-position by $q\sim L_{31}$\/,
anti-momentum by $p=\4P L_{41}$ and the regulator by $r= \4R L_{34}$.
Both the general-quantized $i$ and the general-quantized anti-Hamiltonian 
are linear in $r\sim L_{31}$, with different regulants as coefficients.

This general-quantization necessarily introduces a sixth operator,
 a boost $b\sim L_{12}$
 that vanishes in the q/c limit.
 Relative to the q/c theory, the general-quantized dynamics thus requires two regulators $r, \ b$\/.
To represent the states of the oscillator of ordinary experiments
it suffices to use a representation of high casimir $L^{\alpha}{}_{\beta}L^{\beta}{}_{\alpha}$
($\alpha, \beta = 1, 2, 3, 4, 5$) in a state-vector space $U$ of high dimension.

To represent this dynamical q/q oscillator we may  use a
Clifford algebra  $U=\Cl (Nu)$ 
as state-vector space.

Let us identify the six oscillator operators 
$\8q,\8p, \8t, \8E,  \8r, \8b$ with multiples of
the six infinitesimal generators
$L_{ij}=-L_{ji}\sim x_i\partial_j-x_j\partial_i  $  of $\mbox{SO}(3,1)$ .
These may be represented by operators on a space $u$ with 
unit vectors $\gamma_{1}, \dots, \gamma_{4}$,
and
contravariant metric tensor 
\BE
g^{-1}=-\gamma_{1}\ox \gamma_{1} +\gamma_2\ox \gamma_2 + \gamma_3\ox \gamma_3 + \gamma_4\ox \gamma_4\/,
\EE
and they obey the familiar commutation relations
\BE
L_{ij}\,\4x\, L_{kl}=\frac 12[L_{ik}g_{jl} -L_{il}g_{jk} -L_{jk}g_{il}+L_{jl}g_{ik}].
\EE
Each of these commutation relations has the form of  one  of the three  typical 
forms
\BEA
L_{13}\,\4x\, L_{13} &=&0,\cr
L_{13}\,\4x\, L_{14} &=&L_{34},\cr
L_{24}\,\4x\, L_{13} &=& 0
\EEA
up to a sign,
according to whether the two index pairs $ij$ and $kl$ differ in 0, 1 or 2  indices.

Notation: We designate by $\gamma_{ij}$  six second-grad basis elements of the Clifford algebra $\Cl(3,1)$.
We write $\0L \gamma$ for  left-multiplication with $\gamma$,
$\0R  \gamma$ for right-multiplication with $\gamma$\/,
and $\0D:= \0L-\0R$ for commutation.
All $\0L \alpha$'s commute with all $\0R\beta$'s.

To represent the variables of the q/q event, with their huge spectra, 
we introduce $N$ anti-commuting
replicas $\gamma_{\mu}(n)$ of the above $\gamma_{\mu}$ quartets
($\mu=1,2,3,4; $, $n=1,\dots, N$).
and set
\BE
\label{eq:GAMMA}
\Gamma_{\mu\nu}=\frac 12 \sum_n [\gamma_{\mu}(n), \gamma_{\nu}(n)]\/.
\EE
We use the resulting Clifford algebra 
\BE
\label{eq:u}
U:=\Cl(\mbox{osc}):= \Cl(3N, N)
\EE
 as 
the state-vector space of the q/q event of the general-quantized oscillator,
of dimension $2^{4N}$ before reduction. 
We designate by $\8L_{ij}$ representatives of the same Lie algebra 
acting on $U=\Cl(3N,N)$\/,  the state-vector space  of the q event.

$\mathrm{Endo}\; U$ is  the algebra generated by the $\0L \gamma$'s and the $\0R \gamma$'s.
The physical representation of the $\gamma_{ij}$ on $U$ is
\BE
\gamma_{ij}\to \8L_{\mu\nu}=\0D \Gamma_{\mu\nu}\/.
\EE

Let us introduce six regulants $\4Q_{jk}$ 
as scale factors for the six $\8L_{jk}$, and 
represent the six variables $\8q, \8p, \8t, \8E, \8b, \8r$ as 
$q_{ij}=\4Q_{ij}L_{ij}$ (no summation) 
according to the convention
\BE
\begin{array}{lcrlcrlcrlcr}
\label{eq:CONVENTION}
\8b&=&- q_{12}&:=&-\frac 12 \4Q_{12}\0D L_{12},\quad 
\8q&=& +q_{23}&:=&+\frac 12 \4Q_{23}\0D L_{23}, \cr 
\8p&=&+ q_{24}&:=&+\frac 12\4Q_{24}\0D L_{24}, \quad
\8t&=& -q_{13}&:=&-\frac 12\4Q_{13}\0D L_{13}, \cr 
\8E&=&- q_{14}&:=&-\frac 12\4Q_{14}\0D L_{14},\quad 
\8r&=& +q_{34}&:=&+\frac 12\4Q_{34}\0D L_{34}.
\end{array}
\EE
Then $\4Q_{jk}$ is the quantum unit of the variable $q^{jk}$.
The maximum eigenvalue of $|\8r|$ is  $N\4Q_{34}/l$\/.

It is sometimes helpful to designate the quantum unit of any variable $v$ 
by $\4Q_v$ (for ``the quantum of $v$'').
For example, $\4Q_t=\4Q_{13}=:\4T$ is 
the quantum of time,  and $\4Q_E=\4Q_{14}$ is the quantum of energy.

Since $q_{..}$ and $L_{..}$ are both skew-symmetric,
we may take the matrix $\4Q_{..}$ of quantum cosntants 
to be symmetric.
A more covariant description would relate the 
two second-rank tensors $q_{..}$ and $L_{..}$ 
by a mixed fourth rank tensor $\4Q^{..}_{..}$.
The coefficients $\4Q_{..}$ are eigenvalues of $\4Q^{..}_{..}$.

Written out,  the 15  commutation relations are
\BE
\label{eq:15COMMUTATORS}
\begin{array}{rclrclrcl}
\8q\,\4x\,\8p&=&+\frac 12 \frac{\4Q_q \4Q_p}{\4Q_r} i \8r,&\quad 
\8r\,\4x\,\8q&=&+\frac 12 \frac{\4Q_r \4Q_q}{\4Q_p} i \8p,&\quad 
\8p\,\4x\, \8r&=&+\frac 12 \frac{\4Q_p \4Q_r}{\4Q_q}i \8q, \cr
\8t\,\4x\,\8E&=&+\frac 12 \frac{\4Q_t \4Q_E}{\4Q_r} i \8r, &\quad
\8r\,\4x\, \8t&=&+\frac 12 \frac{\4Q_r \4Q_q}{\4Q_E} i \8E,&\quad
\8q\,\4x\,\8t&=&-\frac 12 \frac{\4Q_q \4Q_t}{\4Q_b}i \8b, \cr
\8E\,\4x\, \8r&=&+\frac 12 \frac{\4Q_p \4Q_r}{\4Q_t} i \8t,&\quad 
\8E\,\4x\, \8q & = &0,&\quad
\8E\,\4x\, \8p & = &+\frac 12 \frac{\4Q_E \4Q_p}{\4Q_b}i \8b,\cr%
\8b\,\4x\, \8E&=&+\frac 12 \frac {\4Q_b \4Q_E}{\4Q_p} i\8p, &\quad %
\8b\,\4x\, \8p&=&-\frac 12 \frac{\4Q_b \4Q_p}{\4Q_E} i \8E,&\quad%
\8b\,\4x\, \8q&=&-\frac 12 \frac{\4Q_b \4Q_q}{\4Q_t} i \8t, %
\cr
\8b\,\4x\, \8t&=& +\frac 12 \frac{\4Q_E \4Q_p}{\4Q_q} i \8q,&\quad%
\8b\,\4x\, \8r&=&0,&\quad %
\8p\,\4x\, \8t&=&0.
\end{array}%OK
\EE
with 15 structure  constants of the form $\hbar_{vw}=\4Q_v \4Q_w/ \4Q_u$
for $L_v \4x L_w= L_u\ne 0$.
These relations  define the general-quantized Lie algebra   $\8L=d \0S\0O(3,1)$.

Each non-zero commutation relation implies a relation between
quantum constants, such as 
\BEA
\hbar & =&\frac {\4Q_q\4Q_p}{\4Q_r}\cr
\hbar  &=&\frac{\4Q_t\4Q_E}{\4Q_r}, \cr
\hbar_{rq}&=&\frac{\4Q_r\4Q_q}{\4Q_p},\cr
&\vdots & 
\EEA
% ={\hbar_{14}\hbar_{34}}/{\hbar_{13}},\cr
% \hbar_{pr}&=&{\hbar_{24}\hbar_{34}}/{\hbar_{23}},\quad
% &
% \hbar_{Ep}&=&{ \hbar_{13}\hbar_{23}}/{\hbar_{12}}
% ={\hbar_{14}\hbar_{24}}/{\hbar_{12}},\cr
%  \hbar_{bE}&=&{\hbar_{12}\hbar_{14}}/{\hbar_{24}}
% 		={\hbar_{12}\hbar_{13}}/{\hbar_{23}},\quad&
% \hbar_{bp}&=&{\hbar_{12}\hbar_{24}}/{\hbar_{14}},\cr
% \hbar_{bq}&=&{\hbar_{12}\hbar_{23}}/{\hbar_{13}},\quad&&& 
% \end{array}
The Jacobi identities
\BE 
\0D[a\, \4x\, b]= [\0D b\,\4x \,\0D a]
\EE
relate the structure constants among themselves.
  For example, 
$\0D t$ acting on $\8q\,\4x\, \8p$ produces
\BEA
\0D\8t\cdot [\8q\, \4x\, \8p] &=& \0D\8t\cdot \hbar i \8r\cr
&=&-\hbar^2\8E\cr
&=& -\0D  [\8q\,\4x\,\8p] \cdot \8t\cr
& = &(\0D \8q \0D\8p -\0D\8p \0D\8q)\cdot \8t\cr
&=& \0D\8p \hbar_{qt} i\8b\/,\cr
&=&\hbar_{pb} \hbar_{qt}E\/,\cr
\hbar^2&=&\hbar_{pb}\hbar_{qt},\cr
\4Q_t \4Q_r^2&=&\4Q_E\4Q_q\4Q_p
\EEA
To be sure,  if we apply $\0D \8r $\/,  $\0D b$\/,  
$\0D p$\/, or $\0D q$  to  this $\8q, \8p$ commutation relation,
we obtain only $0=0$.

The q/q history of the oscillator is a q set of q events.
Its supporting vector space is therefore $U=\Cl u$.

The same relations (\ref{eq:15COMMUTATORS})
hold for the exponentially higher-dimensional Clifford algebra
$V(\0M)=\Cl T=\22^ T$ that  
is the state-vector space of the dynamical history
and supports the q/q dynamical algebra.

\section{Consequences of  general-quantization }

\label{sec:discussion}

The most salient consequence of general-quantization is that it gives all 
variables of the simple system finite bounded discrete spectra,
without reducing the number of continuous symmetries.
On the A line, the basic variables represent generators of an orthogonal group
and  are whole
multiples of a basic quantum unit.

At the same time, the simplicity principle has virtually driven us to a
revised concept of dynamics and time.
A simple dynamics is not defined by  a one-parameter group of unitary 
time translation operators.
The system time is a quantum variable of discrete spectrum
and its advance is not a unitary transformation.
Near the beginning and ending of the system time, 
at $|t|\sim\pm \4N\4T$,
the multiplicities of the eigenvalues of $|t|$ 
vary so rapidly with $t$  --- namely linearly in 
the difference $|t|-\mathrm{max}|t|$ ---
that it is a bad approximation to suppose that the different values of $|t|$ have
isomorphic  eigenspaces related by unitary transformations.
In middle times, $|t|\ll \mathrm{max}|t|$, 
unitarity is a good approximation.
In  this toy model, space becomes small as time nears its beginning or end.
A similar thing happens near the beginning of time in general relativity too,
but it is too soon  to say whether the two phenomena are related.

The relation between the quantities $i$ (a constant) and $\8H$ (a variable) of the old  q/c synchronic dynamics
now resembles that between constant mass and variable energy of the old Newtonian physics.
$\8H$ and $i$ are actually the same variable $r$ seen through different 
lenses.  
The window through which we see $i$ covers the entire range of values of $\8r$,
and in ordinary situations $\8r$ remains so close to its extreme value that 
it can be treated as constant. 
That constant, rescaled to unit magnitude, is $i$.
On the other hand $\8r$ is not exactly constant, and
its departure from its extreme value, again suitably rescaled,
is $\8H$.

Each previous   general-quantization  has introduced new forms of energy with important consequences.
It remains to be seen what  consequences this new concept of energy will have.

Since quantum theory began as a regularization procedure of Planck,
it is rather widely accepted that further regularization of present
quantum physics calls for further quantization, 
but what to quantize
and how to quantize it remains at least a bit unclear. 
Now we regard quantization as a special case of  flexing,
and a path becomes clearer.
It is
marked by singular Killing forms  ripe for  flexing.
All the singular groups and 
infinities of present physics arguably result from  flattening,
and this originates in
a preference for singular groups over regular ones
that is not based on experiment but on ideology.

Flexing the  time-independent 
linear harmonic oscillator  results in a finite quantum theory
with three quantum constants 
$\hbar, \hbar', \hbar''$ instead of the
usual one. This finite quantum oscillator
 is isomorphic to a dipole rotator with
$N=2l+1\sim 1/(\hbar'\hbar'')\gg 1$ states and bounded
Hamiltonian $H= A
(L_1)^2 + B(L_2)^2$.
Its position and momentum variables are 
quantized with uniformly spaced bounded 
finite spectra
and supposedly universal quanta of position and momentum.
For fixed quantum constants and large $N\gg 1$ there
are three broad classes of finite oscillator,
 soft, medium, and hard.
The field
oscillators responsible for infra-red and ultraviolet divergences
are soft and hard respectively. 
Medium oscillators have  $\sim \sqrt N$ low-lying states
having nearly the same zero-point energy and level spacing as the
quantum oscillator and nearly obeying the Heisenberg 
uncertainty principle and
the equipartition principle. 
The corresponding rotators are nearly
polarized along the $z$ axis with $L_3\sim \pm l$. 

The soft and hard
oscillators  have infinitesimal 0-point energy, 
and grossly violate
both equipartition and  the Heisenberg uncertainty relation.
They do not resemble the quantum oscillator at all. 
Their low-lying
energy states correspond to rotators with  $L_1\sim 0 $ or $L_2\sim
0$ instead of  $L_3\sim \pm l$. 
Soft oscillators have frozen
momentum $p\approx 0$ because their maximum potential energy
 is too small to
produce one quantum of momentum. 
Hard oscillators have frozen
position $q\approx 0$  because their maximum kinetic energy 
is too small to
produce one quantum of position.

The zero-point energy of a physical oscillator likely contributes to its
gravitational field.
 It will be interesting to estimate its
contribution to astronomical gravitational fields. 
 For a consistent
estimate  we should regularize the space-time  operators
$x^{\mu}, \pd_{\mu}$ 
as well as the canonical field variables $q, p$, 
since both algebras have
 the same instability. 
 This changes not only the structure of the 
 individual oscillators, as considered here,
 but also the number and distribution of the 
 oscillators.  We leave this study for later.
 
 The finite quantum theory
 modifies low- and high-energy physics.
 Because the low-lying energy levels 
 of medium oscillators have nearly uniform spacing,
 the energy of two excitations is but slightly less
 than the sum of their separate energies.
 The corresponding quanta nearly do not interact,
and the small interaction that they have is attractive.
 For soft or hard oscillators,
 the energy level varies quadratically 
 with the energy quantum number.
 The energy of two quanta of oscillation is twice the sum 
 of their separate energies, for example.
 The corresponding quanta have a repulsive interaction of great strength;
 the interaction energy is equal to the total energy of the separate quanta.
 Thus the simplest regularization leads to
 interactions between the previously uncoupled 
excitation quanta of the oscillator,
weakly attractive for medium quanta, strongly repulsive 
for soft or hard quanta.

Like Dirac's theory of the ``anomalous'' magnetic moment of the 
relativistic electron,
 these extreme-energy effects depend on factor ordering.
They can be adjusted to fit the data by re-ordering factors and so
are not crucial tests of the theory.
The theory of a more physical system will be necessary for that.
 
\section{Acknowledgments and references}

Helpful discussions with James Baugh, Eric Carlen,  
Andrei Galiautdinov, Alex Kuzmich, 
Zbigniew 
Oziewicz, Heinrich Saller, Raphael Sorkin, and John Wood  
are gratefully acknowledged.

\newpage
\tableofcontents

\end{document}